\documentclass[12pt]{article}
\usepackage{setspace}
\usepackage[margin=0.90in]{geometry}
\usepackage{amsmath}
\usepackage{amsfonts}
\usepackage{multicol}
\usepackage{empheq}
\usepackage{caption}
\usepackage{float}
\usepackage{hyperref}
\usepackage{amsrefs} % must be loaded after hyperref
\usepackage{amssymb}
\usepackage{amsthm}
\usepackage{array}
\usepackage{bm}
\usepackage{datetime}
\usepackage{amsmath}
\usepackage{fancyhdr}
\usepackage{fontenc}
\usepackage{ifthen}
\usepackage{accents}
\usepackage{tabularx}

\newtheorem{Theorem}{Theorem}
\newtheorem{Proposition}{Proposition}
\newtheorem{Corollary}{Corollary}

\newcommand{\Aa}{A}
\newcommand{\invform}{\lambda}
\newcommand{\rain}{{\cal R}}
\newcommand{\A}{{\cal A}}
\newcommand{\NC}{\kappa}

\newcommand{\G}{{\bf G} }

\newcommand\barM{
	\hbox{\kern 2.3 true pt 
	\vbox{\hrule width 7  true pt height .3 true pt \kern .9 true pt
	\hbox{\kern -2.3 true pt $M$}}}}

\title{\bf A New Non-Inheriting Homogeneous Solution of the Einstein-Maxwell Equations with Cosmological Term}
\author{ 
Ian Anderson
\\ Department of Mathematics and Statistics
\\ Utah State University  
\\ Logan, Utah 
\\USA 84322
\and 
Charles Torre%\footnote{corresponding author}
\\Department of Physics
\\ Utah State University
\\Logan, Utah 
\\USA 84322
\\
\\
}
\date{January 2022}                                      % Activate to display a given date or no date

\begin{document}
\numberwithin{equation}{section}
\numberwithin{Definition}{section}
\numberwithin{Theorem}{section}
\numberwithin{Proposition}{section}
\numberwithin{Corollary}{section}

\maketitle
 \thispagestyle{empty}
\vskip 0.5truein
%\section{}
%\subsection{}
\begin{abstract}

We find a new homogeneous solution to the Einstein-Maxwell equations with a cosmological term. The spacetime manifold is ${\bf R}\times {\bf S}^3$.  The spacetime  metric  admits a simply transitive isometry group $G = {\bf R} \times {\bf SU(2)}$ and is  Petrov type I. The spacetime is geodesically complete and globally hyperbolic.  The electromagnetic field is non-null and non-inheriting: it is only invariant with respect to the {\bf SU(2)} subgroup and is time-dependent in a stationary reference frame.

\end{abstract}

%\tableofcontents
\vfill\eject
\clearpage
\pagenumbering{arabic} 
\section{Introduction}

In this paper we will find a new homogeneous solution to the Einstein-Maxwell equations with a  cosmological term which has a  non-null and  non-inheriting electromagnetic field. This solution is the only  one known  with all these properties. 

Homogeneous spacetimes -- those admitting a transitive group of isometries -- represent the most symmetric class of spacetime geometries.  This class of spacetimes includes Minkowski and (anti-) de Sitter spacetimes, of course, but other well-known  solutions to the Einstein equations are homogeneous, {\it e.g.} Einstein's static universe, G\"odel's spacetime, the Bertotti-Robinson electrovacuum, and a class of plane waves. The homogeneous spacetimes solving the Einstein equations are all known in a number of cases. See Chapter 12 of reference \cite{Stephani} for a survey of known homogeneous solutions to the Einstein field equations.  

A convenient organization of homogeneous solutions according to the complexity of the field equations is as follows.  The simplest field equations describe spacetimes whose isometry group is multiply transitive and whose matter fields ``inherit'' the full isometry group.
 In the multiply transitive case the isotropy reduces the number of metric and matter components to be considered and the number of field equations is reduced as well. When the matter fields inherit the spacetime symmetries the field equations always reduce to purely algebraic equations. If the isometry group is simply transitive and/or the matter fields do not inherit the spacetime symmetries, the field equations are considerably more complicated. 
Indeed, in the simply transitive case the only known solutions for the classical matter fields are: a vacuum spacetime due to Petrov \cite{Petrov:1962}; an Einstein space due to Kaigorodov \cite{Kaigorodov:1962}; a non-inheriting  electrovacuum  (without cosmological term) due to McLenaghan and Tariq \cite{McLenaghan-Tariq:1975a};  an inheriting Einstein-Maxwell solution with cosmological term due to Ozsvath \cite{Ozsvath:1965a}; and a collection of perfect fluid solutions due to Ozsvath \cite{Ozsvath:1965b}, Farnsworth and Kerr \cite{Farnsworth-Kerr:1966}.   
A complete classification of spacetimes admitting simply transitive isometry groups (due to the authors)  \cite{Anderson-Torre:2020}  shows this list to be complete.

The set of homogeneous solutions to the Einstein-Maxwell equations has been completely determined for null electromagnetic fields including a possible cosmological term, and for inheriting non-null electromagnetic fields without a cosmological term \cite{Stephani}. In reference \cite{Ozsvath} Ozsvath  finds solutions in the inheriting case with a non-zero cosmological term.  We have been able to prove there are no more solutions in that sector.  The non-inheriting case has not been fully explored. Elsewhere we shall show that the non-inheriting solution found by McLenaghan and Tariq is the only non-null, non-inheriting homogeneous solution of the Einstein-Maxwell equations without a cosmological term.  When the cosmological constant is non-zero, we are able to show that the non-null, non-inheriting Einstein-Maxwell solutions (if any) must have a simply transitive isometry group whose corresponding isometry algebra has a derived subalgebra that is simple. So far we are able to find one such solution, which we describe here.  

The solution constitutes a 1-parameter family of spacetimes; the magnitude of the parameter determines the positive cosmological constant.  The spacetime admits the isometry group ${\bf G} = {\bf R}\times {\bf SU(2)}$ acting simply transitively, so the spacetime manifold can be identified with $\bf G$.  There are no hypersurface orthogonal timelike Killing vector fields so the solution is stationary but not static.  The spacetime is Petrov type I; it  is geodesically complete and globally hyperbolic.
 The electromagnetic field for this solution is non-inheriting in that it is invariant only under the {\bf SU(2)} subgroup of the isometry group.  This allows for time-dependent behavior of the electromagnetic field in the setting of the  stationary spacetime.  The electromagnetic invariants are time-dependent and not simultaneously vanishing, so the electromagnetic field is non-null.  In a $\bf G$-invariant reference frame associated to the stationary observers, the electric and magnetic fields trace out ellipses in a plane and the Poynting vector field is along geodesics orthogonal to that plane.
 
  According to the classification results of \cite{Anderson-Torre:2020},  the ``spacetime group'' derived here is in the class denoted by {\bf simpCT}:  spacetimes with a simply transitive isometry group whose isometry Lie algebra has (i) a simple derived algebra ({\bf su(2)}), and (ii)  a center  ({\bf R}) that is a family of  timelike lines. The solution we have found is the only solution to the Einstein-Maxwell equations (with or without a cosmological term) of type   {\bf simpCT}.  More generally, the solution found here can be viewed as a stationary Bianchi IX electromagnetic universe. Bianchi IX  universes with inheriting electromagnetic fields have been studied by Waller \cite{Waller1984} in the case of vanishing cosmological constant. 
  
In the next section we review the Rainich formalism, which characterizes non-null electrovacua (with or without a cosmological term) purely in terms of the metric.  This formalism is particularly convenient for deriving our solution since the metric is homogeneous while the electromagnetic field is not. In \S\ref{SolutionSection} we use the Rainich equations to derive the solution. In \S\ref{PropertiesSection} we display salient mathematical and physical properties of the solution.  An Appendix gives details of the determination of the Petrov type of the spacetime. A worksheet containing details of the computations of \S3--4 can be found at {\tt https://digitalcommons.usu.edu/dg\_applications/}. 

\section{The Einstein-Maxwell equations and their Rainich formulation}
\label{RainichSection}

Let $M$ be a 4-dimensional manifold equipped with a Lorentz metric $g$, and let $F$ be a 2-form on $M$.  We will consider solutions to the Einstein-Maxwell equations in the absence of electromagnetic sources:
\begin{equation}
R_{ab} - \frac{1}{2}R g_{ab} + \Lambda g_{ab} = \NC^2\left(F_{a}{}^c F_{bc} - \frac{1}{4}F_{mn}F^{mn} g_{ab}\right)
\label{EFE}
\end{equation}
\begin{equation}
\nabla_a F^{ab} = 0 = \nabla_{[a}F_{bc]}.
\label{ME}
\end{equation}
Here $\kappa^2$ is the gravitational constant. $R_{ab}$ and $R$ are, respectively, the Ricci tensor and Ricci scalar of $g$. We have generalized the traditional definition of electrovacua to include a cosmological constant $\Lambda$, which may or may not vanish. The derivative $\nabla$ is torsion-free and compatible with the metric.  We only consider solutions in which $F$ is not identically zero.  An electromagnetic field is ``null'' if both of its scalar invariants vanish:
\begin{equation}
F_{ab}F^{ab} = 0 = F_{ab}\tilde F^{ab},
\end{equation}
where $\tilde F_{ab} = \frac{1}{2} \epsilon_{ab}{}^{cd} F_{cd}$.
Otherwise, the electromagnetic field is ``non-null''.

Besides the well-known diffeomorphism symmetry\footnote{Here ``symmetry'' means a transformation of $(g, F)$ which maps solutions of the field equations to other solutions.} admitted by the Einstein-Maxwell equations (\ref{EFE}), (\ref{ME}), there is also an SO(2) group of symmetries defined by duality rotations:
\begin{equation}
F  \to \cos(\phi) F - \sin(\phi) \tilde F, \quad \phi \in [0, 2\pi).
\label{DualityRotation}
\end{equation}

We recall that an isometry of the spacetime $(M,g)$ is a diffeomorphism $\Phi\colon M \to M$ such that $\Phi^*g = g$.  The set of isometries of a given spacetime forms a Lie group, where the group product is composition of diffeomorphisms.
In the next section we will find a family of solutions to (\ref{EFE}), (\ref{ME}) which are {\it homogeneous}, that is, admit a transitive group $\G$ of isometries.  Such solutions will admit at least four   Killing vector fields, four of  which  are point-wise linearly independent.
An electromagnetic field solving (\ref{EFE}), (\ref{ME}) may not admit all the symmetries admitted by the metric. 
In particular,  all that is guaranteed from the covariance of the field equations is that, for each $\Phi\in \G$, there exists a function $\gamma\colon M \to {\bf R}$ such that the electromagnetic 2-form transforms by a duality rotation:
\begin{equation}
\Phi^*F = \cos(\gamma) F - \sin(\gamma) \tilde F.
\end{equation}
If $\gamma = 0$  one says that $F$ inherits that symmetry of the spacetime.  If $F$ inherits all of $\G$, we say that $F$ is ``inheriting''.  Otherwise, $F$ is ``non-inheriting''. 
Infinitesimally, given a  Killing vector field $\xi$ for an Einstein-Maxwell solution, there exists a function $\psi\colon M\to {\bf R}$ such that
\begin{equation}
L_\xi F = \psi \tilde F.
\end{equation}
Here $\psi=0$ if $F$ is inheriting.  Further details about the electromagnetic inheritance of isometries can be found, {\it e.g.}, in references  \cite{McIntosh:1978}, \cite{Henneaux:1984}.

To analyze the field equations we will use the formalism devised by Rainich  \cite{Rainich}. This is a  reformulation of the Einstein-Maxwell equations for a non-null electromagnetic field purely in terms of the metric using the  {\it Rainich equations}:
\begin{equation}
R =4\Lambda, \quad S_a^b S_b^c = \frac{1}{4} \delta_a^c S_{mn}S^{mn} \neq 0, \quad S_{ab}t^at^b >0,
\label{AlgRainNonNull}
\end{equation}
\begin{equation}
\nabla_{[a}\left(\epsilon_{b]cde}{S^c_m\nabla^dS^{me}\over S_{ij}S^{ij}}\right) = 0,
\label{DiffRain}
\end{equation}
where  $S_{ab}$ is the trace-free Ricci tensor, and  $t^a$ is any timelike vector field. These equations generalize Rainich's original results to allow for a cosmological term in the Einstein equations \cite{Krongos-Torre}. 
\medskip
\begin{Theorem}
The Rainich equations (\ref{AlgRainNonNull}), (\ref{DiffRain}) for the metric  $g$ are necessary and sufficient for the existence of a non-null 2-form $F$ such that $(g, F)$ solve the source-free Einstein-Maxwell  equations (\ref{EFE}), (\ref{ME}).
\end{Theorem}

Rainich's formalism also provides an algorithm for constructing the electromagnetic field from a metric satisfying (\ref{AlgRainNonNull}), (\ref{DiffRain}).  The electromagnetic field is non-null if and only if $S_{ab} S^{ab}\neq 0$.
 In particular, given a metric $g$ satisfying the Rainich equations, there exists a 2-form $f$ and a function $\theta$ such that, with
\begin{equation}
F_{ab} = \cos\theta f_{ab} - \sin\theta \tilde f_{ab},
\label{RainichEM}
\end{equation}
$(g, F)$ satisfy the Einstein-Maxwell equations (\ref{EFE}), (\ref{ME}).
The 2-form $f$ is implicitly defined (up to an overall sign) by
\begin{equation}
S_{ab} = \frac{1}{2} \NC^2\left(f_{ac}f_{b}{}^c + \tilde f_{ac} \tilde f_{b}{}^c\right),\quad f_{ab}\tilde f^{ab} =0,\quad f_{ab} f^{ab} < 0,
\label{fdef}
\end{equation}
where
\begin{equation}
\tilde f_{ac} = \frac{1}{2}\epsilon_{ac}{}^{mn} f_{mn}.
\end{equation}
The function $\theta$ is locally determined (up to an additive constant) from the {\it Rainich 1-form}
\begin{equation}
\label{Rainich1form}
\alpha_b := \epsilon_{bcde}{S^c_m\nabla^dS^{me}\over S_{ij}S^{ij}},
\end{equation}
 by
\begin{equation}
\alpha_a =\nabla_a \theta.
\end{equation}
Note that $\alpha$ is closed by virtue of (\ref{DiffRain}).

The 2-form $f_{ab}$ is invariant with respect to the connected component of the isometry group as is the Rainich 1-form.  The function $\theta$ is the only part of the construction of the electromagnetic field that does not depend geometrically and locally on the metric and its derivatives.
 From the perspective of the Rainich theory, the non-inheritance of isometries by the electromagnetic field  stems from the non-invariance of $\theta$ under isometries.  

\begin{Proposition}
\label{prop1}
Given a non-null Einstein-Maxwell solution with Rainich 1-form $\alpha$, the isometry group generated by a Killing vector field $\xi$ is  inherited if and only if 
\begin{equation}
\xi^a \alpha_a = 0.
\end{equation}
\end{Proposition}
\begin{Corollary}
\label{Cor1}
A homogeneous solution of the Einstein-Maxwell equations is inheriting if and only if  the Rainich 1-form vanishes.
\end{Corollary}
\noindent The corollary follows from the fact that  the Killing vector fields span the tangent space at each point of the spacetime manifold $M$ in the case of a transitive action.

We remark that (\ref{RainichEM}) implies the Einstein equations depend upon the electromagnetic field only through its $\G$-invariant part $f_{ab}$, since the energy-momentum tensor is invariant under duality rotations. Only the Maxwell equations depend on the non-{\bf G}-invariant part of the electromagnetic field. 

Finally, it will be helpful in what follows to take account of the behavior of the Einstein-Maxwell equations (or their Rainich form) under a scaling of the metric.
The following result is easily verified.

\begin{Proposition}
\label{ScaleProp}
If $(g_{ab}, F_{ab})$ solve the Einstein-Maxwell equations with cosmological constant $\Lambda$, then $(\hat g_{ab}, \hat F_{ab})$ solve the Einstein-Maxwell equations with cosmological constant $\hat \Lambda$, where
\begin{equation}
\hat g_{ab} = c^2 g_{ab},\quad \hat F_{ab} = c F_{ab},\quad \hat \Lambda = \frac{1}{c^2} \Lambda,
\end{equation}
and $c>0$ is a constant.

\end{Proposition}

\section{The solution}
\label{SolutionSection}

We consider a spacetime $(M, g)$ defined as follows. The spacetime manifold is $M = {\bf R} \times {\bf S}^3$.  The isometry group will be  ${\bf G} = {\bf R} \times {\bf SU(2)}$ with the group ${\bf R}$ acting on ${\bf R}$ by translations (and trivially on ${\bf S}^3$) and ${\bf SU(2)}$ acting on ${\bf S}^3\approx {\bf SU(2)}$ by the canonical left action (and trivially on {\bf R}).  

We will use a basis of 1-forms on ${\bf SU(2)}$ which are invariant with respect to the left action of $\bf SU(2)$ on itself. The are denoted by $(\invform^1, \invform^2, \invform^3)$ and are chosen to satisfy
\begin{equation}
d\invform^1 = \invform^2 \wedge\invform^3,\quad d\invform^2 = \invform^3 \wedge\invform^1,\quad d\invform^3 = \invform^1 \wedge\invform^2. 
\label{MCalg}
\end{equation}
The corresponding dual frame of left-invariant vector fields is denoted by $(l_1, l_2, l_3) $ and satisfies
\begin{equation}
[l_1, l_2] = -l_3,\quad [l_2, l_3]=-l_1,\quad [l_3, l_2] = -l_2.
\end{equation}
A global {\bf G}-invariant co-frame for $M$ is given by
\begin{equation}
\left(dt, \invform^1, \invform^2, \invform^3\right),
\label{coframe}
\end{equation}
and we consider metrics of the form
\begin{equation}
g = -dt \otimes dt + 2 \beta\, dt \odot \invform^1 + a \, \invform^1 \otimes\invform^1 + b\, \invform^2\otimes\invform^2 + c\, \invform^3\otimes\invform^3,
\label{metric}
\end{equation}
where $a$, $b$, $c$, and $\beta$ are constants.  By construction, the 4-parameter family of metrics (\ref{metric}) admits the simply transitive group ${\bf G} =  {\bf R} \times {\bf SU(2)}$ of isometries. 
There is  no loss of generality in using this {\it ansatz}\,: it can be shown  every metric of type  {\bf simpCT}  \cite{Anderson-Torre:2020} which solves the field equations (\ref{EFE}), (\ref{ME}) can be put into the form (\ref{metric}).

The Rainich equations are (\ref{AlgRainNonNull}), (\ref{DiffRain}).  Define 
\begin{equation}
{\cal E}_{ab} = S_a^c S_{cb} - \frac{1}{4} g_{ab} S_{mn}S^{mn},\quad {\cal H}_{ab} = \nabla_{[a}\left(\epsilon_{b]cde}{S^c_m\nabla^dS^{me}\over S_{ij}S^{ij}}\right).
\end{equation}  
The  Rainich equations are equivalent to ${\cal E}_{ab}=0= {\cal H}_{ab}$.  

For the metric (\ref{metric}), the non-trivial Rainich equations are equivalent to
\begin{equation}
\rain_1=\rain_2 = \rain_3 = \rain_4 = \rain_5 = \rain_6 = 0,
\end{equation}
where
\begin{equation}
{\cal R}_1 = \eta{\cal E}_{00},\quad {\cal R}_2 = \eta{\cal E}_{01},\quad {\cal R}_3 = \eta{\cal E}_{11},\quad {\cal R}_4 = \frac{\eta}{b}{\cal E}_{22},\quad {\cal R}_5 = \frac{\eta}{c}{\cal E}_{33},\quad {\cal R}_6 = \zeta{\cal H}_{23}.
\label{R7}
\end{equation}
Here the indices 0, 1, 2, 3 correspond to the coframe (\ref{coframe}), and 
\begin{align}
\eta = 32|\det(g)|^2,\quad \zeta&=16|\det(g)|^{5/2}S_{mn}S^{mn}.
\end{align}

First we show that $\beta\neq0$. Suppose $\beta=0$; this implies that  $a,b,c\neq0$ (or the metric is degenerate). 
With $\beta=0$ it follows that\footnote{This identity was found by using {\it Maple} to construct a Gr\"obner basis for the polynomials ${\cal R}_1, \dots, {\cal R}_6$ in the $\beta=0$ case.  Of course, (\ref{beta0identity}) can be verified directly by  substituting the expressions for ${\cal R}_1, \dots, {\cal R}_6$ (with $\beta=0$)  and then simplifying.}
\begin{equation}
\A_1\rain_1 + \A_4\rain_4 + \A_5 \rain_5 = 768\, b^3 c^4 (b-c)^2,
\label{beta0identity}
\end{equation}
where
\begin{align}
\A_1 &=(b-c)(-4b^4 + 5b^3c - b^2c^2 + 6bc^3 + 34 c^4+3a^2(4b^2 -bc +2c^2))\nonumber\\
&\quad -a(b+c)(8b^3 +2b^2c-15bc^2 + 8c^3)\\
\A_4& = a^2 (8b-7c)(4b^2 - bc + 2c^2)- a(32 b^4-8 b^3 c-4 b^2 c^2-37 b c^3+32 c^4) \nonumber\\
&\quad+c(28 b^4-43 b^3 c+25 b^2 c^2+16 b c^3-2 c^4)\\
\A_5 & =a^2 (7 b-8 c) (4 b^2-b c+2 c^2)  -a(8 b^4+58 b^3 c-61 b^2 c^2+2 b c^3+8 c^4)\nonumber\\
&\quad -( 20 b^5-17 b^4 c-13 b^3 c^2-10 b^2 c^3+20 b c^4-24c^5).
\end{align}
Consequently, the Rainich equations imply $b = c$.  Inserting this into the Rainich equations we find
\begin{equation}
\rain_4 = - a^4,
\end{equation}
but since $a \neq 0$, there can be no solution to the Rainich equations if $\beta=0$. Henceforth we assume $\beta\neq0$.

It is now convenient to introduce a dimensionless time parameter $\tau = t/\beta$ and to express the metric as
\begin{equation}
q = \beta^2\left(-\lambda^0 \otimes \lambda^0 + \Aa\lambda^1 \otimes\lambda^1 + B\lambda^2\otimes \lambda^2 + C\lambda^3 \otimes \lambda^3\right),
\end{equation}
where
\begin{equation}
\lambda^0 = d\tau - \lambda^1,
\end{equation}
and we have defined dimensionless variables $(\Aa, B, C)$:
\begin{equation}
a = (\Aa-1) \beta^2, \quad b = B\beta^2, \quad c = C\beta^2.
\end{equation}
We then have
\begin{align}
&\rain_1 = \beta^8 \left[4\Aa Y - \xi_1 X\right],&
&\rain_2 = \beta^9  \left[-4\Aa Y +\xi_2X\right],&\\
&\rain_3 = \beta^{10}\big[4\Aa(1-\Aa) Y + \xi_3 X\big],&
&\rain_4 = \beta^8\left[4\Aa Y -\xi_4X\right],&\\
&\rain_5 = \beta^8\left[4\Aa Y -  \xi_5 X\right],&
&\rain_6 = 4\beta^{11}(B-C)^2 Z,&
\end{align}
where
\begin{align}
\xi_1&=41\Aa^2 - (46B+14C+9)\Aa+5B^2+6BC+5C^2,\\
\xi_2&=45\Aa^2-(46B+14C+9)\Aa+5B^2+6BC+5C^2,\\
\xi_3&=45\Aa^3 - (46B + 14C + 54)\Aa^2+(B^2+(14C+46)B+C^2+14C+9)\Aa\nonumber \\
&\quad-5B^2-6BC-5C^2,\\
\xi_4 &=  43\Aa^2-(50B + 10 C +7)\Aa+7B^2+10BC-C^2,\\
\xi_5 &= 43\Aa^2 - (42B + 18 C + 7)\Aa-B^2+10BC+7C^2,
\end{align}
and
\begin{align}
X &= \Aa\left[3(\Aa-1) -2(B+C)\right] - (B-C)^2
\label{Xeq}\\
Y &= 32\Aa^3 - (56B+32C+37)\Aa^2+\left[16B^2+(56C+38)B+14C+5\right]\Aa+8B^3\nonumber\\
&\quad-24B^2C-12BC,
\label{Yeq}\\
Z &= \Aa^3-\Aa^2+\left[3B^2+(2C-2)B+3C^2-2C\right]\Aa-4(B+C)(B^2+C^2).
\label{Zeq}
\end{align}

We now show that {\it the Rainich equations are equivalent to the three equations $X=Y=Z=0$.}  To begin, we note that these conditions are sufficient for the Rainich equations to be satisfied.
Next, we observe that $\Aa\neq0$, $B\neq0$, $C\neq0$ or else the metric becomes degenerate. Now we calculate:
\begin{equation}
\beta \rain_1  + \rain_2 = 4\beta^9 \Aa^2 X,
\end{equation}
so that the Rainich equations imply $X=0$.  From, {\it e.g.}, ${\cal R}_1=0$ it then follows that $Y=0$. Finally, if $B = C$ then $\rain_5 = -\beta^8\Aa^2(\Aa-1)^2$, and the Rainich equations will have no solution unless $\Aa=1$. If $B=C$ and $\Aa=1$ , then $X = - 4C$, so that the Rainich equations will have no solution.  Thus we must assume $B\neq C$, and ${\cal R}_6=0$ is then equivalent to $Z=0$.  

We now solve $X=Y=Z=0$ for $\Aa, B, C$. There are two real solutions. 
Introduce new variables
\begin{equation}
u = \frac{1}{3} (B+ C)+\frac{1}{2},\quad  v = -\frac{1}{\sqrt{3}} (B-C),\quad  w = -\frac{1}{3} (B+C)+\Aa -\frac{1}{2}.
\end{equation}
The change of variables yields 
\begin{equation}
X = -3(u^2 + v^2 - w^2).
\end{equation}
Using $X=0$ to eliminate $v^2$ via 
\begin{equation}
v^2= w^2 - u^2,
\label{veq}
\end{equation}
 the equations $Y=0$, $Z=0$ take the form
\begin{align}
\tilde Y &\equiv -16 u^3+(24 w+17) u^2 -(12 w^2+26 w+7) u+w(2 w^2+11 w+\frac{13}{2} )  =0,\\
\tilde Z &\equiv  (2 w-4 u+3) (5 u^2-2 u w+ 2 w^2-8 u+ w+9/4) = 0.
\end{align}
Now we consider the first factor in $\tilde Z$.  If that factor vanishes then it follows from $\tilde Y = 0$ that either $\Aa=0$ or $\Aa=4$.  The former case corresponds to a degenerate metric and the latter case yields a solution with a null trace-free Ricci tensor hence violating (\ref{AlgRainNonNull}).  Thus we can replace $\tilde Z=0$ with 
\begin{equation}
\hat Z \equiv 20 u^2-8 u w+8 w^2-32 u+4 w+9 =0.
\label{hatZ}
\end{equation}
Using this equation to eliminate $w^2$ in $\tilde Y$ results in
\begin{equation}
(2u-1)\left[6u^2 -29u + 15 + (6u+1)w\right] = 0.
\end{equation}
If the first factor vanishes, $u=1/2$, the only solutions to $X=\tilde Y=0$ yield a degenerate metric or $\Aa=-2$, which  gives the wrong signature to the metric.  Consequently, the second factor has to vanish.  It cannot vanish if $u = -1/6$; thus this equation reduces to 
\begin{equation}
w = -\frac{6 u^2-29 u+15}{6 u+1}.
\label{weq}
\end{equation}
If we substitute this result into (\ref{hatZ}) and use $u\neq 1/2$ we get
\begin{equation}
216u^3-756u^2+1170u-583=0,
\end{equation}
which has one real solution:
\begin{equation}
 u = -\frac{2}{3} (2^{1/3} - 2^{-1/3} - \frac{7}{4}).
 \label{ueq}
 \end{equation}
Using (\ref{ueq}), (\ref{weq}), and (\ref{veq}), and converting back to the variables $A, B, C$ gives
\begin{align}
\Aa &= 3 - 2^{1/3},\nonumber\\
B &= 2^{-1/3}-2^{1/3}+1\mp \frac{1}{2}\sqrt{-\,(2^{2/3}) 7+2^{1/3}+10},\nonumber\\ 
 C &= 2^{-1/3}-2^{1/3}+1\pm \frac{1}{2}\sqrt{-\,(2^{2/3}) 7+2^{1/3}+10}.\nonumber\\
 \label{ABCsol}
\end{align}

 In terms of the original variables $a, b, c$ we thus obtain two families of  solutions parametrized by $\beta$:
\begin{align}
a &= \left(2 - 2^{1/3}\right)\beta^2,\nonumber\\
 b &= \left(2^{-1/3}-2^{1/3}+1\mp \frac{1}{2}\sqrt{-\,(2^{2/3}) 7+2^{1/3}+10}\right)\beta^2,\nonumber\\ 
 c &= \left(2^{-1/3}-2^{1/3}+1\pm \frac{1}{2}\sqrt{-\,(2^{2/3}) 7+2^{1/3}+10}\right)\beta^2.\nonumber\\
 \label{TheSolution}
\end{align}
Note that the two sign choices correspond to a change of basis, $\invform^2 \leftrightarrow\invform^3$,  so one can adopt a single sign choice with no loss of generality.  In what follows we shall adopt the upper sign choice. 

We summarize the results of this section as follows.  For each $\beta>0$, the following metric on ${\bf R}\times {\bf S}^3$ satisfies the Rainich conditions (\ref{AlgRainNonNull}) and (\ref{DiffRain}):
\begin{equation}
q  = \beta^2\left(-\lambda^0 \otimes\lambda^0 + \Aa\lambda^1 \otimes \lambda^1 + B\lambda^2 \otimes\lambda^2 + C\lambda^3 \otimes\lambda^3\right).
\label{finalq}
\end{equation}
Here $\Aa$, $B$, $C$ are given in (\ref{ABCsol}), $\lambda^i$, $i=1,2,3$ are invariant forms on ${\bf SU(2)}\approx {\bf S}^3$ satisfying (\ref{MCalg}), and
\begin{equation}
\lambda^0 = d\tau - \lambda^1.
\end{equation}
In light of Proposition \ref{ScaleProp}, we see that the scale set by $\beta$ will determine the cosmological constant and the  scale of the electromagnetic field.

\section{Properties of the solution}
\label{PropertiesSection}

The spacetime defined by (\ref{finalq}) is  Petrov type I (see the Appendix) and has a positive cosmological constant given by
\begin{equation}
\Lambda = \frac{1}{4} R = 2 \left(\frac{2^{2/3} + 2^{1/3} 3 + 9}{25\beta}\right)^2.
\end{equation} 
The $\tau=const.$ hypersurfaces can be identified with the Lie group {\bf SU(2)} equipped with a left-invariant Riemannian metric, and so each is geodesically complete. It follows, {\it e.g.,} from the results of reference \cite{Cotsakis}, that the spacetime is globally hyperbolic and geodesically complete.

The electromagnetic field for this solution is non-inheriting.  To see this, we compute the Rainich 1-form (\ref{Rainich1form}):
\begin{equation}
\alpha = \frac{\sqrt{2}}{2}\left(2  + 2^\frac{2}{3} + 2^\frac{1}{3}\right)\,d\tau\equiv \omega\, d\tau.
\label{Rainich1formSol}
\end{equation}
Because the 1-form has a non-vanishing $d\tau$ component, the electromagnetic field associated to this solution will not be invariant under translations in $\tau$ (see Proposition \ref{prop1} and Corollary \ref{Cor1}).

The following is a {\bf G}-invariant orthonormal frame of  vector fields and its  dual co-frame,
\begin{align}
e_0 =\frac{1}{\beta} \partial_\tau,\quad e_1 = \frac{1}{\beta\sqrt{\Aa}}\left( \partial_\tau + l_1\right), \quad 
e_2=\frac{1}{\beta\sqrt{B}} l_2,\quad e_3=\frac{1}{\beta\sqrt{C}} l_3,
\label{GOT}
\end{align}
\begin{equation} 
\theta^0 = \beta(d\tau - \invform^1)=\beta\lambda^0,\quad \theta^1 = \beta\sqrt{\Aa}\invform^1,\quad\theta^2= \beta\sqrt{B}\invform^2,\quad \theta^3 = \beta\sqrt{C}\invform^3.
\end{equation}
We remark that  each of the  vector fields in (\ref{GOT}) is tangent to a congruence of geodesics.  In terms of the orthonormal co-frame,  
the electromagnetic field is, up to a duality rotation, given by 
\begin{equation}
\NC\,F = \sigma\, \sin(\omega \tau) \,\theta^0 \wedge \theta^2
+ \rho\, \cos(\omega \tau)\, \theta^0 \wedge \theta^3
+ \rho\,  \sin(\omega \tau) \,\theta^1\wedge\theta^2
+ \sigma\, \cos(\omega \tau) \,\theta^1\wedge\theta^3,
\end{equation}
where\begin{align}
\sigma & =\frac{1}{25\beta} \sqrt{\nu -\mu},\\
\rho & = \frac{1}{25\beta}\sqrt{\nu + \mu},
\end{align}
and
\begin{equation}
\mu = \frac{1}{{5}}\sqrt{(-7)2^{\frac{2}{3}}+2^{\frac{1}{3}}+10}\left(738 + (557) 2^\frac{2}{3} + (646) 2^\frac{1}{3}\right),
\end{equation}
\begin{equation}
\nu = 592 + (313) 2^\frac{2}{3} + (464) 2^\frac{1}{3}.
\end{equation}
The electromagnetic invariants are given by
\begin{align}
\frac{1}{2} F_{ab}F^{ab} &= -\frac{2\mu}{625\kappa^2\beta^2}\cos(2\omega \tau),
\label{I1} \\
\frac{1}{4}\epsilon^{abcd}F_{ab}F_{cd} &= -\frac{2\mu}{625\kappa^2\beta^2}\sin(2\omega \tau).
\label{I2}
\end{align}

An  observer with 4-velocity $e_0$  is freely falling (geodesic) and ``sees'' the spacetime as stationary.\footnote{There are no static families of observers, {\it i.e.} hypersurface-orthogonal timelike Killing vector fields.}  The geodesic congruence of such observers has vanishing expansion and shear, but non-vanishing rotation 2-form ${\bf \Omega}$, given by
\begin{equation}
{\bf \Omega}
 = -\frac{\beta}{2} \lambda^2\wedge \lambda^3.
\end{equation}
According to the stationary observer, the vectors $e_2$ and $e_3$ in (\ref{GOT}) are rotating about  $e_1$  with angular frequency $\Omega$:
\begin{equation}
\nabla_{e_0} e_1 = 0,\quad \nabla_{e_0} e_2 = \Omega e_3,\quad \nabla_{e_0} e_3 = - \Omega e_2,
\end{equation}
where
\begin{equation}
\Omega^2 \equiv \frac{1}{2}\Omega^{ab}\Omega_{ab} = \frac{203 + (259)2^{-\frac{1}{3}} + (176) 2^\frac{1}{3}}{625\beta^2}.
\end{equation}
In the $\bf G$-invariant orthonormal frame (\ref{GOT}) the electric and magnetic fields each trace out an ellipse in the $e_2$, $e_3$ plane with angular speed $\omega$ determined by the Rainich 1-form (\ref{Rainich1formSol}).  The Poynting vector is constant in time and is directed along $e_1$. 

 A non-rotating orthonormal frame for the stationary observers is given by:
\begin{equation}
f_0 =e_0,\quad f_1 = e_1,\quad
f_2 =  \cos(\Omega t)e_2 - \sin(\Omega t)e_3,\quad
f_3 = \sin(\Omega t)e_2 +  \cos(\Omega t)e_3,
\label{fframe}
\end{equation}
and satisfies
\begin{equation}
\nabla_{f_0} f_\mu = 0,\quad \mu =0,1,2,3 .
\end{equation}
In this (non-{\bf G}-invariant)  frame the time evolution  of the electric and magnetic field vectors is rather more complicated: it involves a superposition of the elliptical motion at frequency $\omega$  with precession at angular speed $\Omega$.

Because the electromagnetic field is non-null, there is a preferred class of observers who ``see'' the electric and magnetic fields as collinear.  A family of such observers, carrying a $\bf G$-invariant orthonormal frame, corresponds to the following  tetrad
\begin{equation}
\left[\frac{1}{\sqrt{\rho^2 - \sigma^2}}(\rho e_0 - \sigma e_1),\frac{1}{\sqrt{\rho^2 - \sigma^2}}( -\sigma e_0 + \rho e_1), e_2, e_3\right].
\end{equation}
In this  frame the electric and magnetic fields both point in the $e_3$ direction with sinusoidally varying amplitudes. The world lines of these (freely falling) observers do not correspond to the integral curves of a timelike Killing vector field so these observers see a time-varying geometry.

\section*{Acknowledgments}
This work was supported in part by National Science Foundation grant ACI-1642404.  The bulk of the computations were performed with the assistance of the {\it DifferentialGeometry} software package \cite{DG}.

%Data sharing not applicable to this article as no datasets were generated or analysed during the current study.

\appendix
\section{Petrov type}

In this appendix we give the details of the determination of the Petrov type of the solution.

According to the results, {\it e.g.}, in references \cite{Stephani}, \cite{Acevedo-Bonilla}, in order to prove that the spacetime investigated here has Petrov type I it is sufficient to establish that 
\begin{equation}
I^3 - 27 J^2 \neq 0,
\end{equation}
where the scalar invariants $I$ and $J$ are given in terms of the Newman-Penrose Weyl scalars by
\begin{align}
I &=  \psi_0\psi_4 - 4\psi_1\psi_3 + 3\psi_2^2\\
J &= \psi_0\left(\psi_2\psi_4 - \psi_3^2\right) + \psi_2\left(2\psi_1\psi_3 - \psi_2^2\right) - \psi_4\psi_1^2.
\end{align}
Using the null tetrad constructed from the vector fields in (\ref{GOT}) according to
\begin{equation}
k = \frac{1}{\sqrt{2}}(e_0 + e_1),\quad l =  \frac{1}{\sqrt{2}}(e_0 - e_1),\quad  m = \frac{1}{\sqrt{2}}(e_2 + i e_3),\quad  \overline m = \frac{1}{\sqrt{2}}(e_2 - i e_3),
\end{equation}
the Weyl scalars are given by
\begin{align}
\psi_0 &= \frac{(B - C)}{4\Aa BC\beta^2}\left[B+C-\Aa-\sqrt{\Aa}\right],\\
\psi_1&=0,\\
\psi_2&=\frac{1}{12\Aa B C \beta^2}
\left[2\Aa^2 - (B+C+2)\Aa - (B-C)^2\right],
 \\
\psi_3&=0,\\
\psi_4 &= \frac{(B - C)}{4\Aa BC\beta^2}\left[B+C-\Aa+\sqrt{\Aa}\right].
\end{align}
Using (\ref{ABCsol}) we obtain
\begin{equation}
I^3 - 27 J^2 =- 4\frac{ (187804027) (3^2) (3557) 2^{2/3} + (369518691691) (41) 2^{-2/3} + (1051)(9071602009)}{5^{24}\beta^{12}}\neq 0.
\end{equation}
\end{document}